\documentclass[apj,square,numbers]{emulateapj}
\usepackage{graphicx}
\usepackage{xcolor}
\usepackage{amsmath,amssymb,latexsym}
\usepackage{hyperref}
\usepackage{ulem}

\begin{document}
\def\gsim{ \lower .75ex \hbox{$\sim$} \llap{\raise .27ex \hbox{$>$}} }
\def\lsim{ \lower .75ex \hbox{$\sim$} \llap{\raise .27ex \hbox{$<$}} }
\title{Curvature from strong gravitational lensing: a spatially closed Universe or systematics?}
\author{Zhengxiang Li$^{1,\ast}$, Xuheng Ding$^{1,2}$, Guo-Jian Wang$^{1}$, Kai Liao$^{3,\dagger}$, and Zong-Hong Zhu$^{1}$}

\affil{$^{1}$Department of Astronomy, Beijing Normal University, Beijing 100875, China; zxli918@bnu.edu.cn}
\affil{$^2$Department of Physics and Astronomy, University of California, Los Angeles, CA, 90095-1547, USA}
\affil{$^{3}$School of Science, Wuhan University of Technology, Wuhan 430070, China; liaokai@mail.bnu.edu.cn}

\begin{abstract}
  Model-independent constraints on the spatial curvature are not only closely related to important problems such as the evolution of the Universe and properties of dark energy, but also provide a test of the validity of the fundamental Copernican principle. In this paper, with the distance sum rule in the Friedmann-Lema\^{i}tre-Robertson-Walker metric, we achieve model-independent measurements of the spatial curvature from the latest type Ia supernovae and strong gravitational lensing (SGL) observations. We find that a spatially closed Universe is preferred. Moreover, by considering different kinds of velocity dispersion and subsample, we study possible factors which might affect model-independent estimations for the spatial curvature from SGL observations. It is suggested that the combination of observational data from different surveys might cause a systematic bias and the tension between the spatially flat Universe and SGL observations is alleviated when the subsample only from the Sloan Lens ACS Survey is used or a more complex treatment for the density profile of lenses is considered. 
\end{abstract}
\maketitle
\renewcommand{\baselinestretch}{1.5}

\section{Introduction}
The spatial property of the Universe is one of the issues which are at the root of cosmology. Specifically, a particularly important assumption that the Universe, on average, is exactly described by the homogeneous and isotropic Friedmann-Lema\^{i}tre-Robertson-Walker (FLRW) metric, have played a pivotal role from the beginning of the modern history of the subject~\citep{Ellis2006}. This fundamental assumption is also called as the Copernican principle (CP). However, possibilities for the failure of the FLRW approximation have been proposed to account for the observed late-time accelerated expansion~\citep{Ferrer2006, Ferrer2009, Enqvist2008, Redlich2014, Rasanen2009, Lavinto2013, Boehm2013}. Furthermore, even if the FLRW metric is valid, whether the space of the Universe is open, flat, or closed is one of the most fundamental problems in modern cosmology because the curvature of the Universe is closely related to the evolution of the Universe and the nature of dark energy. For instance, on the one hand, any significant deviation from the flat case would lead to profound consequences for inflation models and fundamental physics. On the other hand, nonzero curvature may result in enormous effects on reconstructing the state equation of dark energy even though the true curvature might be very small~\citep{Ichikawa2006, Ichikawa2007, Clarkson2007, Gong2007, Virey2008}. Because of the strong degeneracy between the curvature of the Universe and the dark energy equation of state, it is difficult to constrain these two parameters simultaneously. Therefore, the curvature is usually left out in dark energy studies, or conversely, a constant dark energy equation of state is assumed for determining the curvature. Up to now, constraints on the cosmic curvature from popular observational probes have been widely investigated in the literature~\citep{Eisenstein2005, Tegmark2006, Zhao2007, Wright2007}. Most notably, in the framework of the standard $\Lambda$CDM model, a spatially flat Universe is favored at very high confidence level by the latest {\it Planck} 2015 results of Cosmic Microwave Background (CMB) observations~\citep{Planck2015}. However, it should be stressed here that all these works have not measured the curvature in any
direct geometrical way. That is, the curvature is primarily derived from measurements which are not only dependent on curvature but also on the choice of cosmological model assumed in the analysis.

In addition to obtaining tight constraints on cosmological parameters in specific models, there is also a growing realization that we have to test the fundamental assumptions of our cosmological models as rigorously as we can. The increasing precision and breadth of cosmological observations makes it possible to test assumptions behind entire classes of models. Recently, \citet{Clarkson2008} proposed to directly measure the spatial curvature of the Universe at different redshifts or even test the radial homogeneity in a model-independent way by combining observations of the expansion rate and distance. This null test has been fully implemented with updated observational data~\citep{Shafieloo2010, Mortsell2011, Sapone2014, Yulong2014, Ronggen2016}. These tests consistently suggested that there is no significant deviation of the FLRW metric. In addition, another kind of important by-products of these tests is model-independent estimations for the spatial curvature. However, in this method, derivative of distance with respect to redshift $z$ introduces a considerable uncertainty in estimating the curvature. Therefore, constraints on the curvature with greater precision from observations of expansion rate and distance have been recently obtained by dodging the derivative of distance with respect to redshift~\citep{Yuhai2016, Zhengxiang2016, Junjie2016}. Meanwhile, a similar test has also been put forward to check the validity of the FLRW models by using parallax distances and angular diameter
distances~\citep{Rasanen2014}. More recently, the sum rule of distances along null geodesics of the FLRW metric has been put forth as a consistency test~\citep{Rasanen2015}. It is interesting to note that, on the one hand, the FLRW background will be ruled out if the distance sum rule is violated; on the other hand, if the observational data are well consistent with the distance sum rule, the test provides a model-independent estimation for the spatial curvature of the Universe. In fact, the distance sum rule has already been proposed to be a practical measurement of the curvature of the Universe by studying the cross-correlation between foreground mass and gravitational shear of background galaxies~\citep{Bernstein2006}. In \citet{Rasanen2015}, by using
the Union2.1 compilation of type Ia supernova (SNe Ia)~\citep{Suzuki2012} and strong gravitational lensing (SGL) data selected from the Sloan Lens ACS Survey (SLACS)~\citep{Bolton2008}, the spatial curvature parameter is constrained to be $\Omega_k=-0.55_{-0.67}^{+1.18}$ at $95\%$ confidence level, which slightly favors a spatially closed Universe.

In this paper, following the method proposed in \citet{Rasanen2015}, we update constraints on the spatial curvature by confronting the latest joint light-curve analysis (JLA) SNe Ia~\citep{Betoule2014} with the largest compilation of SGL observations~\citep{Cao2015}. Firstly, for the full sample including all SGL systems, it is suggested that a closed Universe is preferred at more than 95\% confidence level. That is to say, the flat case is
hardly compatible with current observations. Furthermore, we investigate the influence of possible factors in SGL observations on the estimation of cosmic curvature. Specifically, we take the properties of images, the masses of deflectors (corresponding to the dispersion velocities), the combination of data from different surveys, and the reduced $\chi^2$ of SGL sample, into consideration. We find that the inconsistency between the spatially flat Universe and SGL observations might be slightly relieved when the subsample only from the SLACS is used.

This paper is organized as follows. In Sec. \ref{sec2}, we describe the general equations used for our analysis. In Sec. \ref{sec3}, we present the observational data and corresponding constraint results. Finally, conclusions and discussions are presented in Sec. \ref{sec4}.

\section{Methodology}\label{sec2}
In a homogeneous and isotropic Universe with maximum symmetry, the spacetime is described by the FLRW metric (in units where $c=1$)
\begin{equation}
  \textrm{d}s^2=-\textrm{d}t^2+a^2(t)\left(\frac{\textrm{d}r^2}
  {1-Kr^2}+r^2\textrm{d}\Omega^2\right),
\end{equation}
where $a(t)$ is the scale factor and $K$ is a constant relating to the geometry of three dimensional space. Let $d_\mathrm{A}(z_l,z_s)$ represent the angular diameter distance of a source at redshift $z_s$ as observed at redshift $z_l$, then the dimensionless comoving angular diameter distance $d(z_l,z_s)\equiv(1+z_s)H_0d_A(z_l,z_s)$
can be written as
\begin{equation}
  d(z_l, z_s)=\frac{1}{\sqrt{\mid\Omega_k\mid}}S_K\bigg(\sqrt{\mid\Omega_k\mid}
  \int_{z_l}^{z_s}\frac{\textrm{d}x}{E(x)}\bigg),
\end{equation}
where
\begin{equation}\label{dzlzs}
  S_K(X)=\left\{
  \begin{array}{ll}
    \displaystyle \sin(X) &~~~~~\Omega_k<0 \\
    \displaystyle X &~~~~~\Omega_k=0 \\
    \displaystyle \sinh(X) &~~~~~\Omega_k>0
  \end{array}\right.
\end{equation}
$\Omega_k\equiv-K/H_0^2a_0^2$ ($a_0=a(0)$ and $H_0$ are the present values of the scale factor and the Hubble parameter $H=\dot{a}/a$,
respectively), and $E(z)\equiv H(z)/H_0$. In addition, we respectively denote $d(z)\equiv d(0,z)$, $d_l\equiv d(0,z_l)$, $d_s\equiv d(0,z_s)$, and $d_{ls}\equiv d(z_l,z_s)$. If the relation between the cosmic time $t$ and redshift $z$ is a single-valued function and $d'(z)>0$, these distances in the FLRW frame are connected via a simple sum rule~\citep{Peebles1993}
\begin{equation}\label{sum1}
  d_{ls}=d_s\sqrt{1+\Omega_kd_l^2}-d_l\sqrt{1+\Omega_kd_s^2}.
\end{equation}
Apparently, the distances can be simply added together in a spatially flat Universe. Note that, there is $d_s>d_l+d_{ls}$ or $d_s<d_l+d_{ls}$ for $\Omega_k>0$ or $\Omega_k<0$, respectively (see Figure 1 in \citet{Bernstein2006} for an illustration). Chronologically, the fundamental sum rule of Eq.~(\ref{sum1}) was first proposed to obtain model-independent estimate of the spatial curvature by combining weak lensing with baryon acoustic oscillations (BAO) measurements~\citep{Bernstein2006}. Furthermore, Eq.~(\ref{sum1}) can be rewritten as
\begin{equation}\label{sum2}
  \frac{d_{ls}}{d_s}=\sqrt{1+\Omega_kd_l^2}-\frac{d_l}{d_s}\sqrt{1+\Omega_kd_s^2}.
\end{equation}
More recently, on the basis of Eq.~(\ref{sum2}), a model-independent consistency test for the FLRW metric was discussed in \citet{Rasanen2015}, by confronting the distance ratios $d_{ls}/d_s$ derived from the measured SGL systems with distances from SNe Ia observations. It is noteworthy that any inconsistency between the sum rule and observations might imply a deviation from the FLRW metric. In addition, if the observational data are well compatible with the sum rule, the test provides a model-independent measurement for the spatial curvature of the Universe. In this work, following this route, we present an updated estimate of the spatial curvature or even a test of the FLRW metric from the latest JLA SNe Ia and the largest SGL samples.

\section{Data and Results}\label{sec3}
\subsection{Type Ia Supernovae--distances $d_l$ and $d_s$}
In order to get model-independent estimates of the spatial curvature via the simple sum rule, we use the latest JLA SNe Ia to provide distances $d_l$ and $d_s$ in the right-hand-side of the Eq.~(\ref{sum2}). In practice, the distance modulus $\mu$, relating to the luminosity distance $D_\mathrm{L}$ via
$\mu=5\log\big[\frac{D_\mathrm{L}}{\mathrm{Mpc}}\big]+25$, can be directly determined from observed light curves of SNe Ia,
\begin{equation}\label{eq2}
\mu^{\mathrm{SN}}(\alpha, \beta,
M_\mathrm{B})=m_\mathrm{B}^*-M_\mathrm{B}+\alpha\times
x_1-\beta\times c.
\end{equation}
where $\alpha$ and $\beta$ are nuisance parameters which characterize the stretch-luminosity and color-luminosity relationships, reflecting the well-known broader-brighter and bluer-brighter relationships, respectively. The value of $M_\mathrm{B}$ is another nuisance parameter which represents the absolute magnitude of a fiducial SNe. It was found to be dependent on the properties of host galaxies, e.g., the host stellar mass ($M_{\mathrm{stellar}}$). In the latest JLA SNe Ia, this dependence is approximately corrected with a simple step function when the mechanism is not fully understood~\citep{Sullivan2011,Conley2011},
\begin{eqnarray}\label{eq3}
M_\mathrm{B}=\begin{cases} M_\mathrm{B}^1~~~&\mathrm{if}~
M_{\mathrm{stellar}}<10^{10} M_\odot.\\
M_\mathrm{B}^1+\Delta_\mathrm{M}~~&\mathrm{otherwise}.
\end{cases}
\end{eqnarray}

With the distance-duality relation which holds in any spacetime~\citep{Etherington1933, Ellis2009}, the dimensionless comoving angular diameter distance $d=H_0D_\mathrm{L}/(1+z)$ can be obtained by normalizing $H_0$ (here, the latest local measurement $H_0=73.24\pm1.74$~\citep{Riess2016} is used). In principle, we have to select the observed SNe Ia at the certain redshifts which exactly match those of source and lens in SGL systems. Unfortunately, it is impossible to be achieved for all discrete observed events. For this issue, many methods have been proposed to reduce the systematic
uncertainty resulted from the redshift difference between two kinds of observations~\citep{Cardone2012, Liang2013, Holanda2013}. In our analysis, as in  \citet{Rasanen2015}, we model-independently determine the function of dimensionless angular diameter distance with respect to redshift (i.e., $d(z)$) by fitting a polynomial to the JLA SNe Ia data. This function enables us to match all observed SGL systems with redshifts of the source and deflector located in
the range $0<z\leq1.3$ (the maximum redshift of JLA SNe Ia). In this work, we use a simple third-order polynomial function with initial conditions, $d(0)=0$ and $d'(0)=1$, to fit the cosmology-free but light-curve fitting parameters-dependent distances of SNe Ia. This polynomial is expressed as,
\begin{equation}
d(z)=z+a_1z^2+a_2z^3,
\end{equation}
where $a_i$ are two free parameters which need to be constrained simultaneously with light-curve fitting parameters. It has been suggested that, with current data, it dose not make significant difference which function is used, as long as it is more flexible than a second order polynomial~\citep{Rasanen2015}.

\subsection{Strong Gravitational Lensing--distance ratios $d_{ls}/d_s$}
For the distance ratios $d_{ls}/d_s$ in the left-hand-side of Eq.~(\ref{sum2}), they are determined from the measurements for angular separation between strongly lensed images as well as velocity dispersion of the deflector. If the general relativity is valid on the scales of lensing systems and the mass distribution
profile of lenses can be approximately described by a singular isothermal ellipsoid (SIE), the distance ratio can be expressed as
\begin{equation}\label{lens}
\frac{d_{ls}}{d_s}=\frac{\theta_\mathrm{E}}{4\pi f^2\sigma^2},
\end{equation}
where $\theta_\mathrm{E}$ is the Einstein radius, $\sigma$ is the velocity dispersion of the lens and $f$ is a phenomenological coefficient which characterizes uncertainties resulted from the difference between the observed stellar velocity dispersion and that from the SIE model, as well as other systematic effects~\citep{Cao2012}. It should be noted that, in principle, $f$ is strictly equal to 1 when the mass distribution profile of lens galaxies is described by the singular isothermal sphere (SIS) model. Here, the treatment of $f$ is an intractable issue since it significantly degenerates with $\Omega_k$ and the uncertainties in modeling SGL systems is dominant when estimating the spatial curvature. In general, observations suggest the range $0.8<f^2<1.2$~\citep{Kochanek2000, Ofek2003}. However, any prior for $f$, e.g., fixing it at 1 or assigning an extra Gaussian error to it, might lead to bias in the estimations of $\Omega_k$ because of the strong degeneracy between them. Therefore, in our analysis, rather than fixing $f$ at 1 or assigning an extra Gaussian error of $20\%$ to $f^2$ in Ref.~\citep{Rasanen2015}, we take $f$ as a free parameter on the same weight as $\Omega_k$. Following \citet{Bolton2008}, we assign an error of 2\% on $\theta_\mathrm{E}$ and a minimum error of 5\% on $\sigma$. Moreover, two different kinds of velocity dispersion, the velocity dispersion measured within an entire aperture ($\sigma_{\mathrm{ap}}$) and the one measured in a circular aperture of half the effective radius ($\sigma_0$), are often referred in the literature. In theory, $\sigma_{\mathrm{ap}}$ is used for a single system while $\sigma_0$ is applied when we deal with a sample of lenses. Here, we consider both cases. In addition to the case where the lens is approximately described by a singular isothermal ellipsoid (SIE) with one free parameter $f$, we also consider a more complicated SGL model introduced in \citet{Schwab2010},  where Eq.~(\ref{lens}) is replaced by $d_{ls}/d_s=N(\eta, \delta, \epsilon)(\theta_{\mathrm{ap}}/ \theta_{\mathrm{E}})^{\eta-2}/(4\pi\sigma^2)$, with $\eta$, $\delta$, $\epsilon$ and $\theta_{\mathrm{ap}}$ being the slope of the density, anisotropy of the velocity dispersion, the luminosity, and the spectrometer aperture radius, respectively. In this study, $\eta$, $\delta$, and $\epsilon$ are treated as universal parameters. 

In our analysis, SGL systems are selected from the latest compilation presented in \citet{Cao2015}, which includes 118 well-measured galactic-scale lenses from four surveys: the Sloan Lens ACS Survey, the BOSS Emission-Line Lens Survey, the Lenses Structure and Dynamics Survey, and the Strong Lensing Legacy Survey. Due to the limitation of the maximum redshift in the JLA SNe Ia, the maximum source redshift should be cut off at $z=1.3$ and this leaves
us 79 SGL systems. Besides the full 79-events sample, we also consider several subsamples to examine possible influence on the estimation of curvature from factors in SGL observations which might be a source of systematics. According to the property of images in lens, we select systems with two images and those with approximate Einstein ring. The mass density profiles for lens galaxies in these systems are most probably symmetric and thus can be characterized by a single parameter $f$. This subsample is labeled as lensing\_images and consists of 65 lensing systems. As suggested in \citet{Cao2016,Xia2016}, the stellar mass of lens galaxy might lead to possible bias in implications from SGL observations. Therefore, following \citet{Cao2016,Xia2016}, we select systems with typical velocity dispersion of the lens galaxy ranging from 200 to 300 $\mathrm{km}~\mathrm{s}^{-1}$. This subsample is named as lensing\_midmass and includes 55 lensing systems. Moreover, the combination of observations from different surveys also might result in systematic bias~\citep{Xia2016}. Thus, we collect 57 events from the SLACS survey and label this subsample as lensing\_SLACS. Studies on cosmological implications from SGL observations indicate that lensing data usually give a bit large value of $\chi^2$ for per degree of freedom ($\chi^2/\mathrm{d.o.f}$, also named as the reduced $\chi^2$)~\citep{Cao2012,Cao2015,Xia2016}. This implies that there might be some other systematics which are not included. In \citet{Xia2016}, they introduced an extra $\sigma_{\mathrm{int}}$ to represent any other unknown uncertainties except for the observational statistical ones. In their analysis, although the reduced $\chi^2$ remarkably decreased to be close to unit, the constraint results were not significantly changed by the introduction of this extra term. Here, rather than introducing the extra $\sigma_{\mathrm{int}}$, we sift out those points with the corresponding $\chi^2$ greater than 2 to avoid an unreasonable $\chi^2$. We denote this subsample as lensing\_chi2 and it contains 64 events. In short, we summarized these (sub)samples in Table (\ref{Tab1}).

\begin{table*}
	\centering
\begin{tabular}{|c|c|c|}
\hline ~Notation ~&~~Criterion~~&~~Number of events~~\\
\hline
lensing\_full~~&$\diagdown$&~~74~~\\
\hline
lensing\_images~~&~~two images or Einstein rings~~&~~65~~\\
\hline
lensing\_midmass~~&~~$200~\mathrm{km}~\mathrm{s}^{-1}<\sigma_{\mathrm{ap}}
(\sigma_0)<300~\mathrm{km}~\mathrm{s}^{-1}$~~&~~55~~\\
\hline
lensing\_SLACS~~&~~only from the SLACS survey~~&~~57~~\\
\hline
lensing\_chi2~~&~~corresponding $\chi^2<2$~~&~~64~~\\
\hline
\end{tabular}
\tabcolsep 0pt \caption{\label{Tab1} Summary for (sub)samples of lensing systems used in our analysis.} \vspace{0.2cm}
\end{table*}

\subsection{Data fit and Results}
In this work, we infer the value of $\Omega_k$ via Eq. (\ref{sum2}) by confronting measurements of SNe Ia and SGL observations. For the SNe Ia, there are light-curve fitting parameters ($\alpha$, $\beta$, $M_\mathrm{B}^1$, and $\Delta_\mathrm{M}$) accounting for distance estimation in SNe Ia observations. For the SGL, there is the phenomenological coefficient ($f$ or $\eta, \delta, \epsilon$) accounting for the estimation of $d_{ls}/d_s$. Together with the polynomial coefficients ($a_1$ and $a_2$), there are, in total, eight free parameters which should be simultaneously constrained from the SNe Ia and SGL datasets:
\begin{equation}
\mathbf{P}=\{\Omega_k, f(\mathrm{or}~ \eta, \delta, \epsilon), \alpha, \beta, M_\mathrm{B}^1,
\Delta_\mathrm{M}, a_1, a_2\}.
\end{equation}
In our analysis, we perform a global fitting with the emcee\footnote{https://pypi.python.org/pypi/emcee} which is introduced by Foreman-Mackey {\it et al.}~\citep{Foreman2012} using the Python module including Markov chain Monte Carlo (MCMC). For the error analysis of JLA SNe Ia, the full covariance matrix propagated from statistical and systematic uncertainties is used (see \citet{Betoule2014, Junjie2016} for details).

\begin{figure*}[h]
\centering
\includegraphics[width=0.45\textwidth, height=0.45\textwidth]{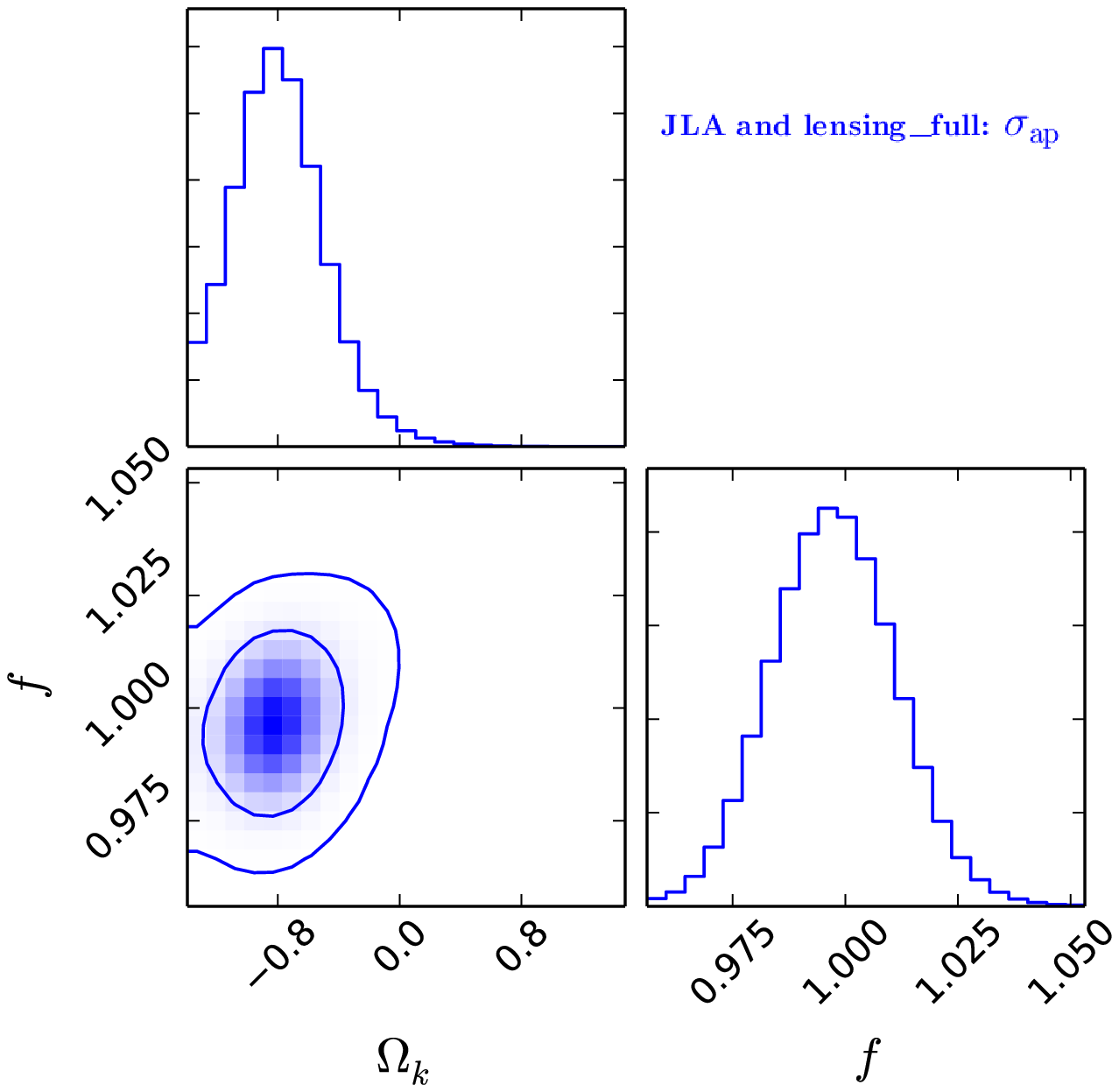}
\includegraphics[width=0.45\textwidth, height=0.45\textwidth]{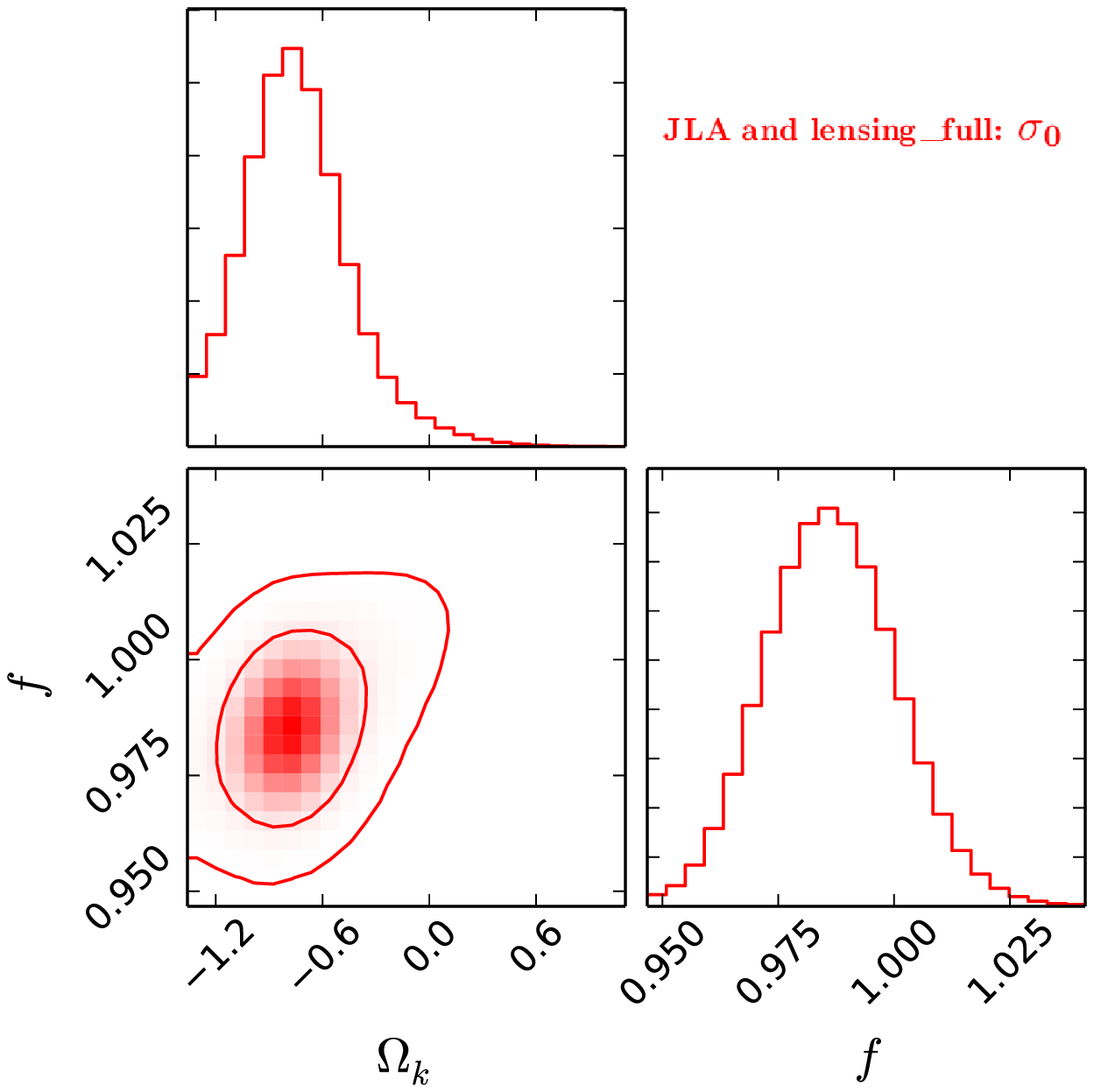}\\
\caption{\label{fig1} The 1-D and 2-D marginalized distributions with 1$\sigma$ and 2$\sigma$ contours for the parameters $\Omega_k$ and $f$ constrained from the JLA SNe Ia and the lensing\_full samples. Results shown in the left and right panels are derived when $\sigma_{\mathrm{ap}}$ and $\sigma_0$ are used, respectively.}
\end{figure*}

\begin{figure*}[h]
		\centering
  \includegraphics[width=0.45\textwidth, height=0.45\textwidth]{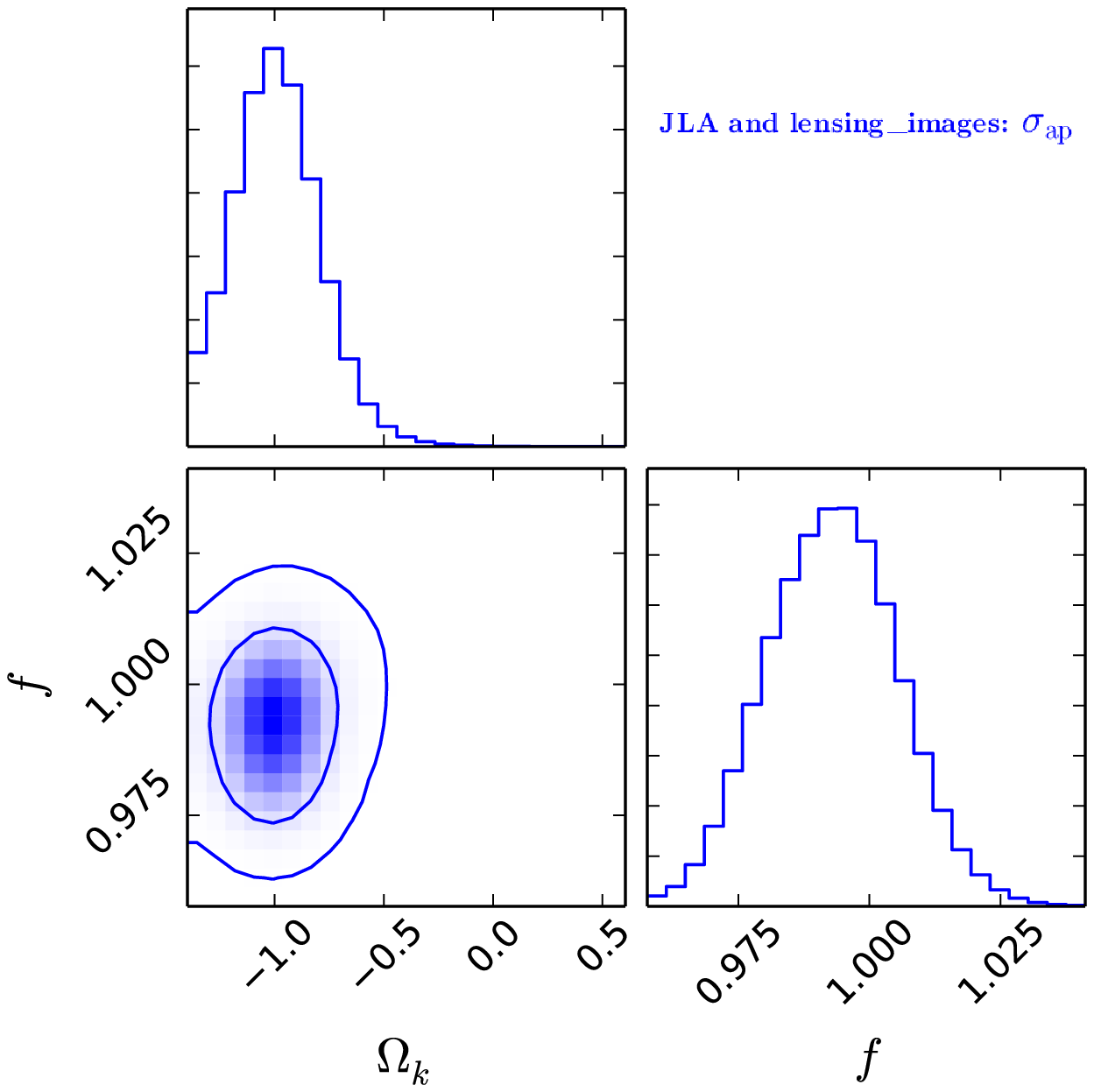}
  \includegraphics[width=0.45\textwidth, height=0.45\textwidth]{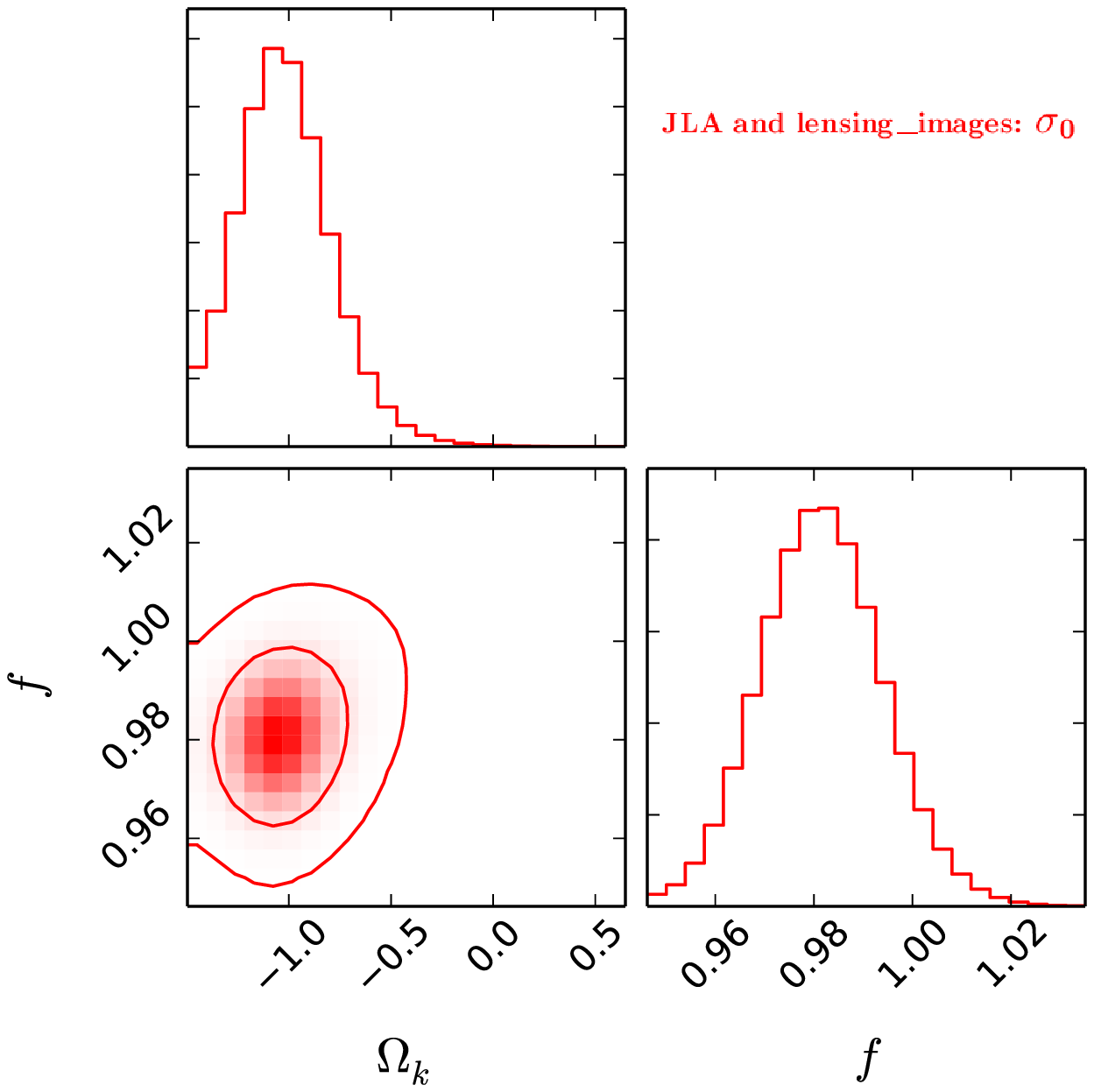}
  \caption{\label{fig2} The same as Figure 1, except now using the lensing\_images subsample. }
\end{figure*}

\begin{figure*}[h]
		\centering
  \includegraphics[width=0.45\textwidth, height=0.45\textwidth]{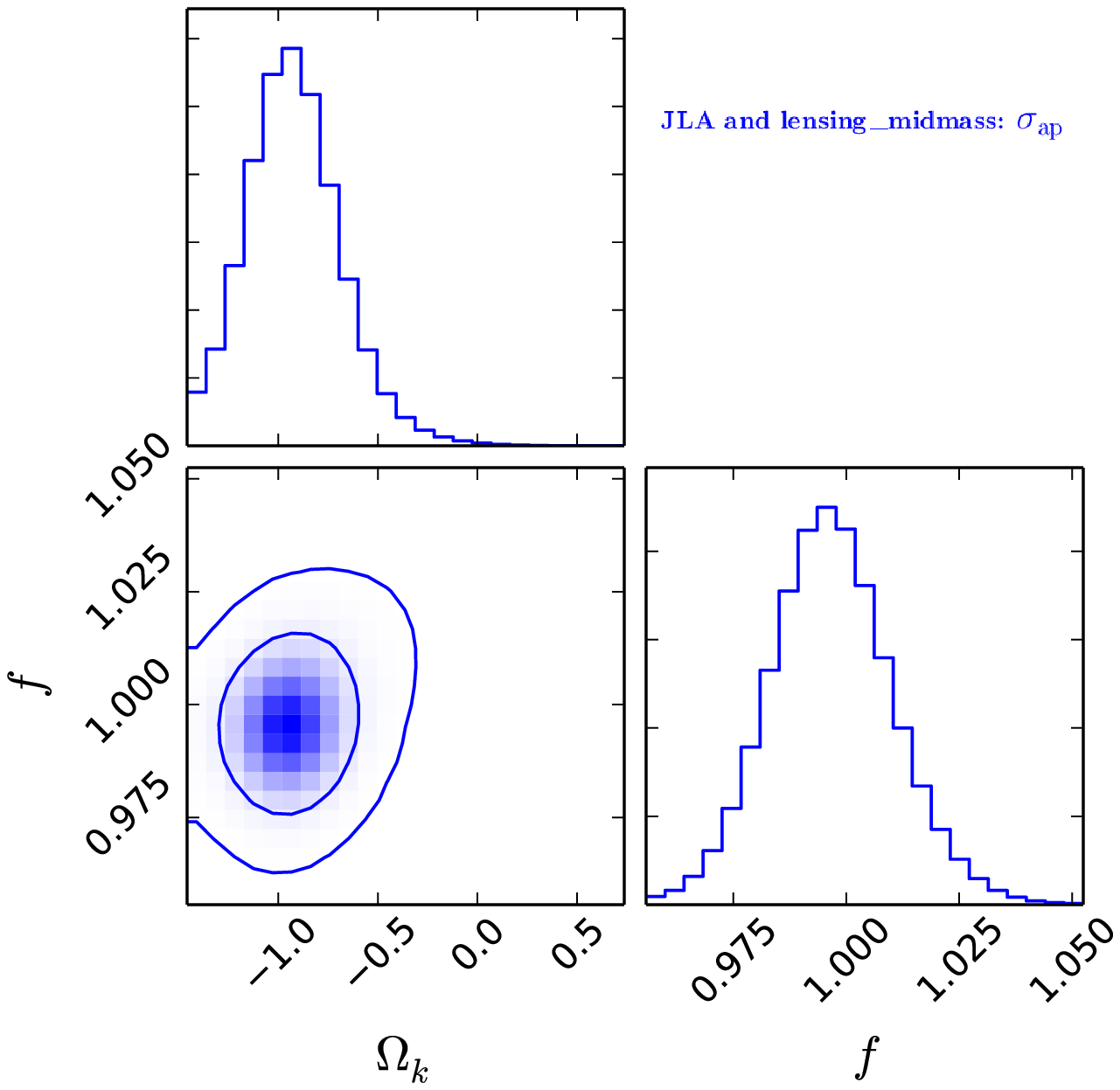}
  \includegraphics[width=0.45\textwidth, height=0.45\textwidth]{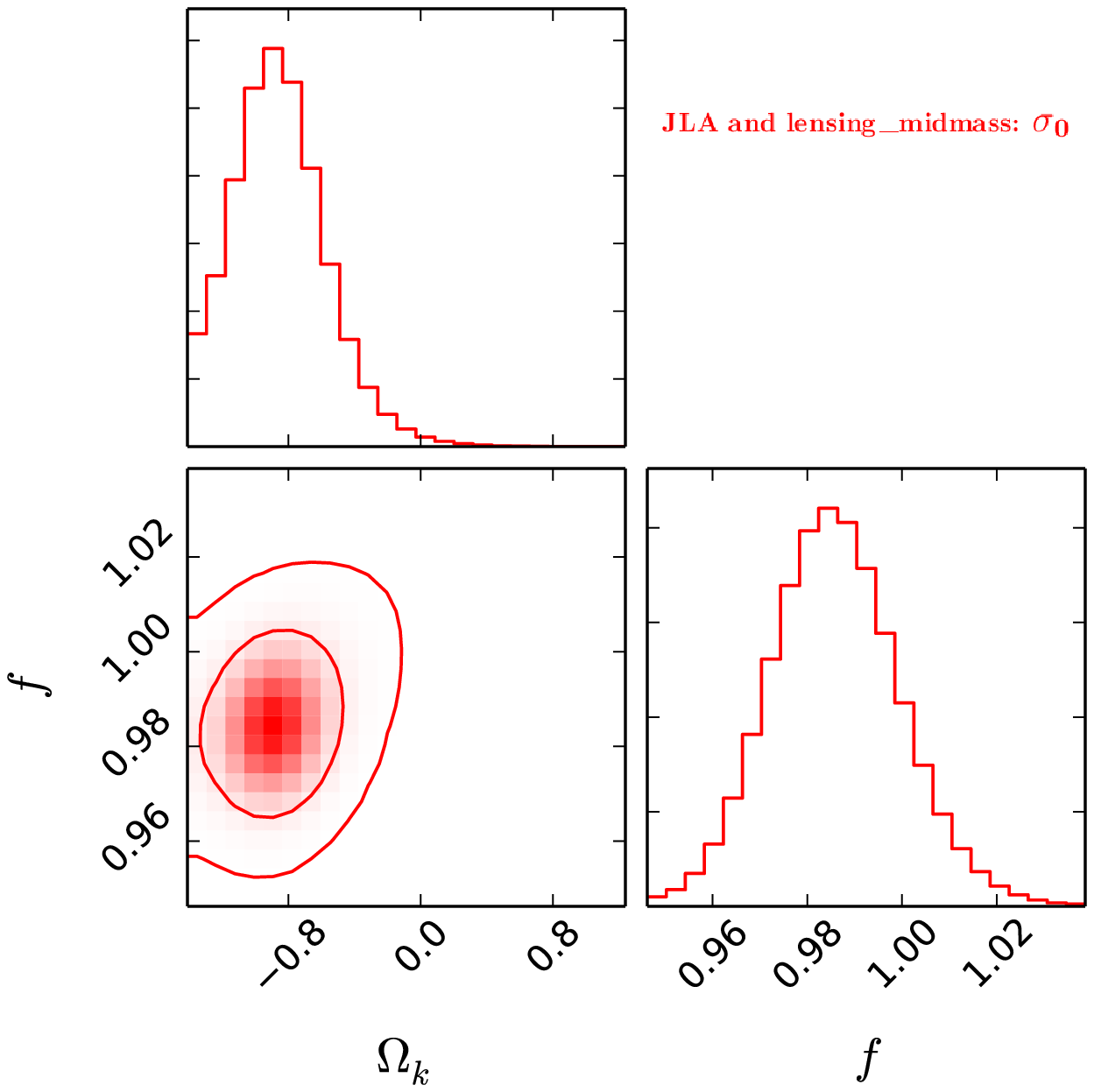}
  \caption{\label{fig3} The same as Figure 1, except now using the lensing\_midmass subsample.}
\end{figure*}

\begin{figure*}[h]
		\centering
  \includegraphics[width=0.45\textwidth, height=0.45\textwidth]{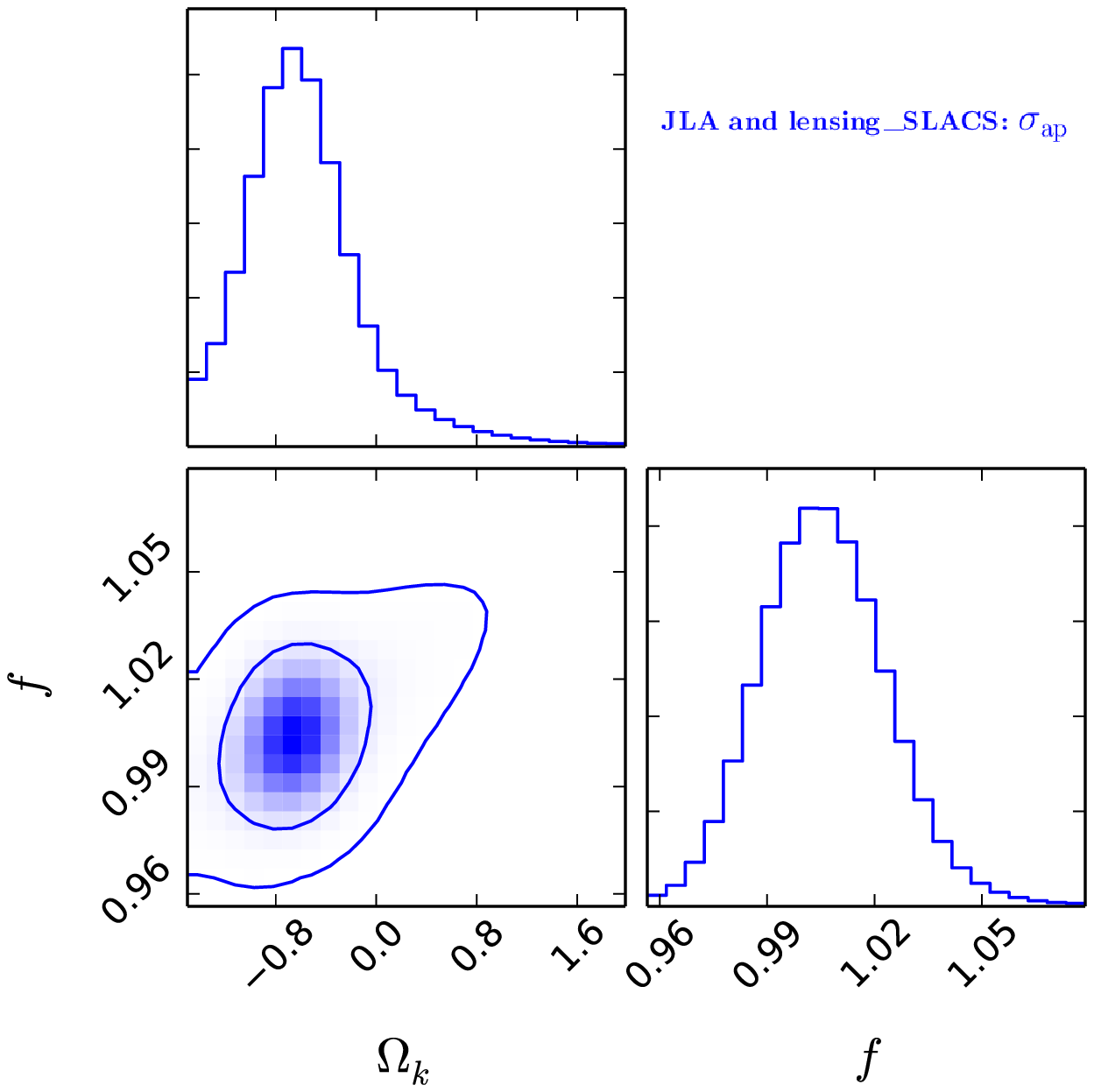}
  \includegraphics[width=0.45\textwidth, height=0.45\textwidth]{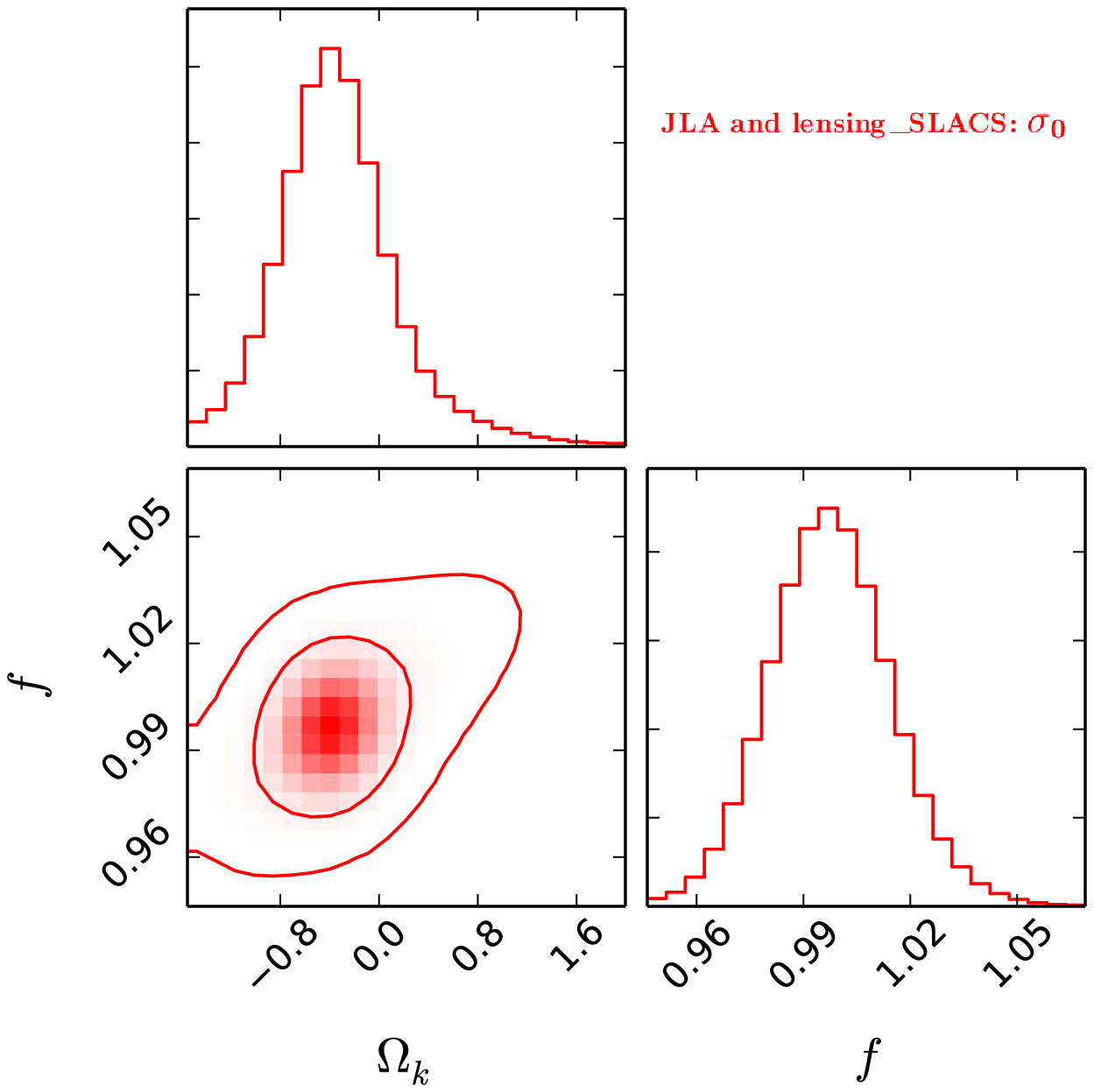}
  \caption{\label{fig4} The same as Figure 1, except now using the lensing\_SLACS subsample.}
\end{figure*}

\begin{figure*}[h]
		\centering
  \includegraphics[width=0.45\textwidth, height=0.45\textwidth]{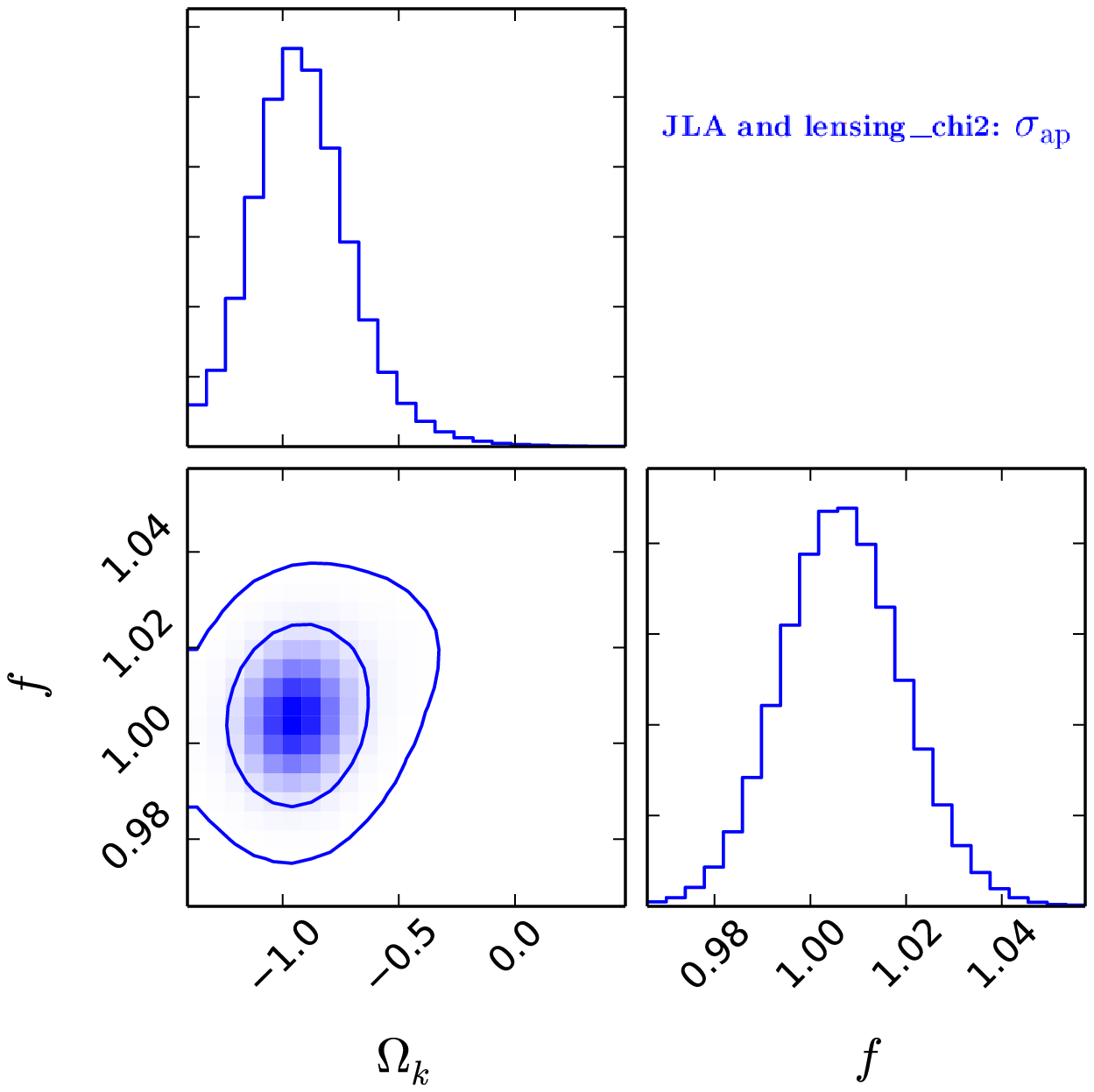}
  \includegraphics[width=0.45\textwidth, height=0.45\textwidth]{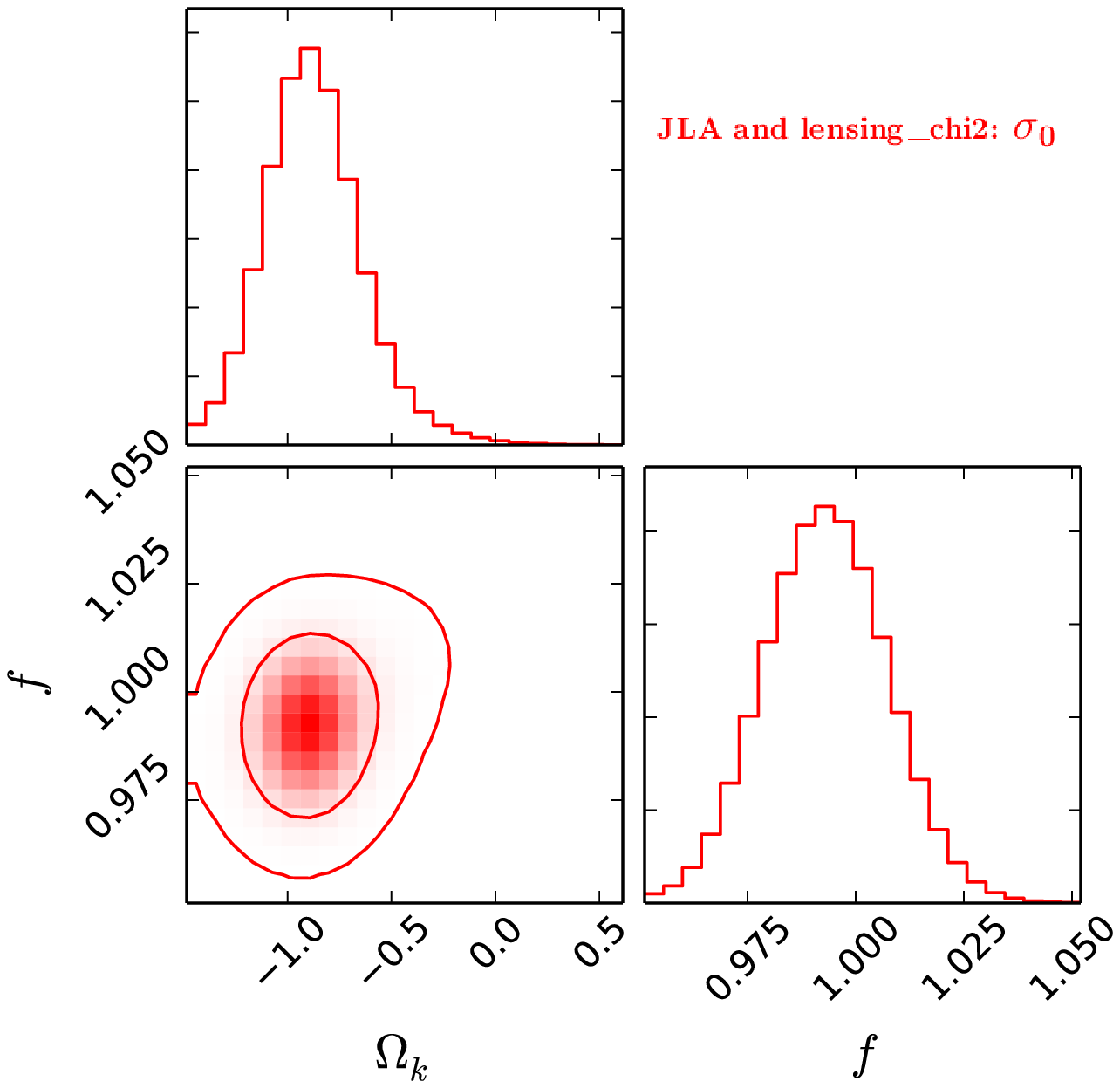}
  \caption{\label{fig5}The same as Figure 1, except now using the lensing\_chi2 subsample.}
\end{figure*}

By marginalizing the light-curve fitting parameters and the polynomial coefficients, we obtain the 1-D and 2-D marginalized distributions with 1$\sigma$ and 2$\sigma$ contours for the parameters $\Omega_k$ and $f$ constrained from the JLA SNe Ia and SGL systems for the simple SIE model. The results are shown in Figures~(\ref{fig1}-\ref{fig5}). Meanwhile, numerical results of constraints on all parameters are summarized in Table~\ref{tab1}. Importantly, compared to the results shown in \citet{Rasanen2015}, constraints on the $\Omega_k$ have been approximately improved by a factor of 5 due to the increase of the number of well-measured lensing systems. In our analysis, we estimate all free parameters in a global fitting without taking any priors for both $\Omega_k$ and $f$ into consideration. However, it should be stressed that estimations for the spatial curvature consistently favor a closed universe at a very high confidence level. In fact, such a trend has already been slightly indicated in \citet{Rasanen2015}. More seriously, due to tighter constraints on $\Omega_k$, the flat case is almost ruled out by our model-independent estimations, which is significantly different from the conclusion of the latest CMB observations~\citep{Planck2015}. For these measurements based on geometrical optics, the tension between the constraints on $\Omega_k$ and the spatially flat Universe is only alleviated when the lensing\_SLACS subsample is used. It may be inferred that the combination of SGL observations from different surveys is the main source of systematics on estimations for $f$, and thus for curvature $\Omega_k$. In addition, the results regarding $f$ suggest that the mass distribution profiles of lens galaxies are statistically well consistent with the simplest SIS model ($f=1$). These results are also remarkably different from what obtained in \citet{Xia2016}. In their analysis, the SIS profile is almost disfavored at more than 95\% confidence level when the prior for the curvature from the $Planck$ 2015 CMB observations ($\Omega_k>-0.1$) is considered. This discrepancy may support our previous claim that any priors for $\Omega_k$ might lead to bias on estimations for $f$ because of the strong degeneracy between them.  Moreover, for the complicated SGL model, numerical results of constraints on all parameters are summarized in Table~\ref{tab2}. It is found that, although the discrepancy between the spatially flat  case and model-independent estimations from SGL observations has eased off because of weak constraints on concerned parameters due to the extension of the lens model, a spatially closed Universe is still slightly favored. 

\begin{figure*}[t]
	\centering
  \includegraphics[width=0.45\textwidth, height=0.45\textwidth]{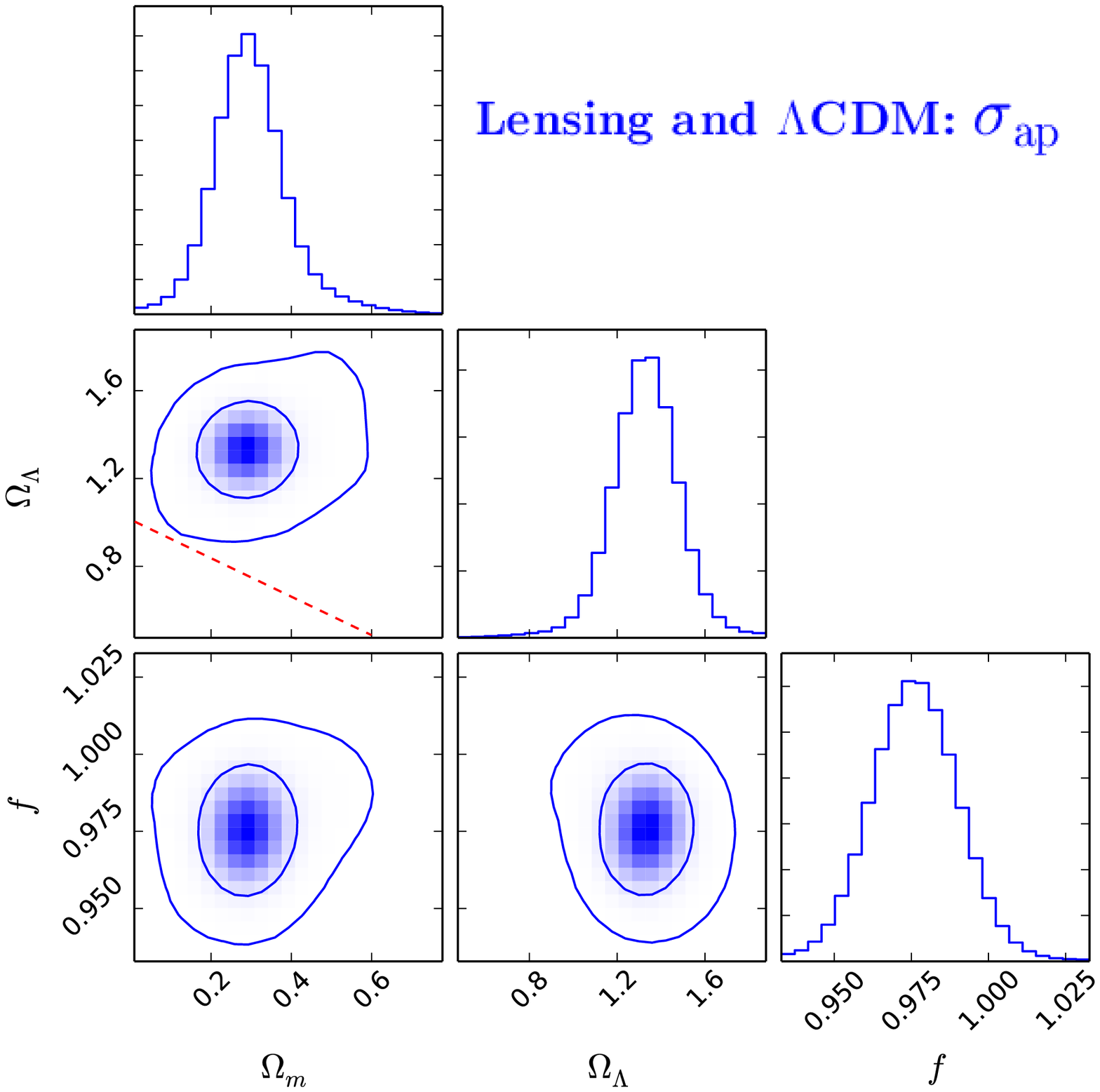}
  \includegraphics[width=0.45\textwidth, height=0.45\textwidth]{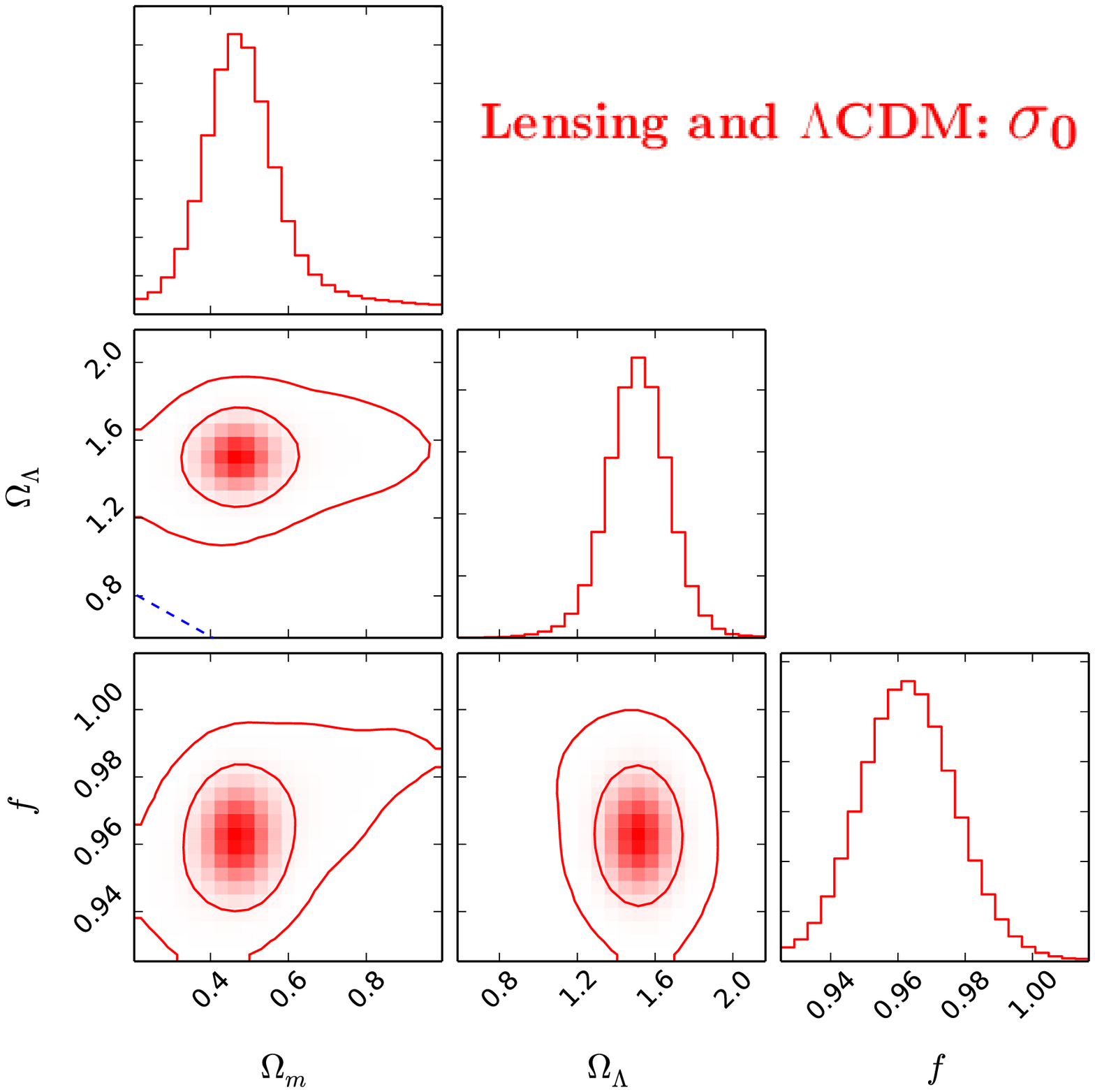}
  \caption{\label{fig6}The 1-D and 2-D marginalized distributions with 1$\sigma$ and 2$\sigma$ contours for the parameters $\Omega_m$, $\Omega_\Lambda$ (in the standard $\Lambda$CDM model), and $f$ constrained from the total 118 SGL systems compiled in \citet{Cao2015}. The left and right panels are results when $\sigma_{\mathrm{ap}}$ and $\sigma_0$ are used, respectively. The blue and red dashed lines denote the spatially flat Universe, $\Omega_m+\Omega_\Lambda=1$. The regions above these lines correspond to the spatially closed Universe.}
\end{figure*}

For comparison, we also investigate constraints on the standard $\Lambda$CDM model from the whole 118 SGL systems. The results are shown in Figure~\ref{fig6}. From these model-dependent estimations, we obtain $\Omega_k=-0.620_{-0.053}^{+0.049}$ and $\Omega_k=-0.986_{-0.041}^{+0.083}$ when $\sigma_{\mathrm{ap}}$ and $\sigma_0$ is used, respectively. It is implied that, a closed Universe is still favored by SGL observations alone at a very high confidence level. Let's keep in mind that, as mentioned in \citet{Rasanen2015}, model-independent estimations for the spatial curvature based on the distance sum rule are also mainly determined by SGL systems. Therefore, both direct geometrical estimations and model-dependent constraints for the curvature from SGL observations consistently favor a spatially closed Universe.

\begin{table*}[t]
	\centering
  \begin{tabular}{cccccc}
    \hline\hline
    Parameters~&~lensing\_full~&~lensing\_images~&~lensing\_midmass~&~lensing\_SLACS~&~
    lensing\_chi2~\\
    \hline
    ~~&~~&&$\sigma_{\mathrm{ap}}$&~~&~~\\
    \hline
    $\Omega_k$~&~$-0.832^{+0.165}_{-0.078}$~&~$-1.029^{+0.094}_{-0.020}$~&
    ~$-0.942^{+0.122}_{-0.072}$~&~$-0.660^{+0.327}_{-0.132}$~&~$-0.947^{+0.144}_{-0.052}$~\\

    $f$~&~$0.997^{+0.011}_{-0.011}$~&~$0.994^{+0.008}_{-0.012}$~&
    ~$0.996^{+0.011}_{-0.009}$~&~$1.005^{+0.015}_{-0.013}$~&~$1.006^{+0.010}_{-0.009}$~\\

    $\alpha$~&~$0.136^{+0.007}_{-0.007}$~&~$0.135^{+0.007}_{-0.009}$~&
    ~$0.137^{+0.007}_{-0.007}$~&~$0.134^{+0.008}_{-0.008}$~&~$0.136^{+0.010}_{-0.012}$~\\

    $\beta$~&~$2.904^{+0.064}_{-0.065}$~&~$2.889^{+0.040}_{-0.039}$~&
    ~$2.902^{+0.052}_{-0.051}$~&~$2.888^{+0.069}_{-0.054}$~&~$2.899^{+0.062}_{-0.061}$~\\

    $M_\mathrm{B}^1$~&~$-18.935^{+0.024}_{-0.023}$~&~$-18.941^{+0.018}_{-0.015}$~&
    ~$-18.942^{+0.033}_{-0.031}$~&~$-18.946^{+0.030}_{-0.019}$~&~$-18.938^{+0.027}_{-0.027}$~\\

    $\Delta_\mathrm{M}$~&~$-0.061^{+0.023}_{-0.024}$~&~$-0.037^{+0.008}_{-0.031}$~&
    ~$-0.063^{+0.023}_{-0.024}$~&~$-0.043^{+0.036}_{-0.035}$~&~$-0.060^{+0.024}_{-0.026}$~\\

    $a_1$~&~$-0.300^{+0.030}_{-0.030}$~&~$-0.311^{+0.026}_{-0.020}$~&
    ~$-0.277^{+0.032}_{-0.030}$~&~$-0.299^{+0.041}_{-0.036}$~&~$-0.294^{+0.037}_{-0.035}$~\\

    $a_2$~&~$0.074^{+0.028}_{-0.026}$~&~$0.089^{+0.009}_{-0.033}$~&
    ~$0.052^{+0.026}_{-0.027}$~&~$0.071^{+0.035}_{-0.039}$~&~$0.070^{+0.037}_{-0.035}$~\\
    
    \hline
    ~~&~~&&$\sigma_0$&~~&~~\\
    \hline
    $\Omega_k$~&~$-0.790^{+0.203}_{-0.108}$~&~$-1.079^{+0.178}_{-0.031}$~&
    ~$-0.913^{+0.185}_{-0.067}$~&~$-0.362^{+0.264}_{-0.208}$~&~$-0.920^{+0.149}_{-0.052}$~\\

    $f$~&~$0.986^{+0.011}_{-0.011}$~&~$0.982^{+0.009}_{-0.009}$~&
    ~$0.985^{+0.011}_{-0.010}$~&~$0.997^{+0.013}_{-0.012}$~&~$0.993^{+0.011}_{-0.011}$~\\

    $\alpha$~&~$0.136^{+0.009}_{-0.009}$~&~$0.137^{+0.010}_{-0.010}$~&
    ~$0.136^{+0.009}_{-0.010}$~&~$0.136^{+0.009}_{-0.010}$~&~$0.136^{+0.010}_{-0.010}$~\\

    $\beta$~&~$2.897^{+0.062}_{-0.065}$~&~$2.912^{+0.072}_{-0.078}$~&
    ~$2.900^{+0.045}_{-0.056}$~&~$2.891^{+0.073}_{-0.056}$~&~$2.900^{+0.062}_{-0.062}$~\\

    $M_\mathrm{B}^1$~&~$-18.936^{+0.022}_{-0.023}$~&~$-18.937^{+0.022}_{-0.023}$~&
    ~$-18.945^{+0.022}_{-0.022}$~&~$-18.940^{+0.024}_{-0.023}$~&~$-18.940^{+0.021}_{-0.021}$~\\

    $\Delta_\mathrm{M}$~&~$-0.061^{+0.023}_{-0.019}$~&~$-0.064^{+0.026}_{-0.026}$~&
    ~$-0.061^{+0.023}_{-0.022}$~&~$-0.060^{+0.024}_{-0.025}$~&~$-0.060^{+0.023}_{-0.024}$~\\

    $a_1$~&~$-0.299^{+0.032}_{-0.034}$~&~$-0.280^{+0.024}_{-0.038}$~&
    ~$-0.279^{+0.030}_{-0.032}$~&~$-0.292^{+0.032}_{-0.032}$~&~$-0.292^{+0.033}_{-0.029}$~\\

    $a_2$~&~$0.074^{+0.035}_{-0.032}$~&~$0.051^{+0.038}_{-0.024}$~&
    ~$0.056^{+0.032}_{-0.030}$~&~$0.067^{+0.037}_{-0.035}$~&~$0.068^{+0.030}_{-0.032}$~\\

    \hline\hline
  \end{tabular}
  \caption{\label{tab1}Constraints on all parameters from the JLA SNe Ia and SGL observations when the SIE model is considered.}\vspace{0.2cm}
\end{table*}

\begin{table*}[t]
    \centering
	\begin{tabular}{cccccc}	
		\hline\hline
		Parameters~&~lensing\_full~&~lensing\_images~&~lensing\_midmass~&~lensing\_SLACS~&~
		lensing\_chi2~\\
		\hline
		~~&~~&&$\sigma_{\mathrm{ap}}$&~~&~~\\
		\hline
		$\Omega_k$~&~$-0.308^{+0.613}_{-0.321}$~&~$-0.662^{+0.370}_{-0.445}$~&
		~$-0.529^{+1.007}_{-0.528}$~&~$-0.353^{+0.543}_{-0.823}$~&~$-0.387^{+1.004}_{-0.699}$~\\
		
		$\eta$~&~$1.981^{+0.314}_{-0.284}$~&~$1.789^{+0.121}_{-0.124}$~&
		~$1.738^{+0.201}_{-0.321}$~&~$1.857^{+0.401}_{-0.355}$~&~$1.973^{+0.386}_{-0.217}$~\\
		
	    $\delta$~&~$1.330^{+0.237}_{-0.541}$~&~$1.463^{+0.312}_{-0.132}$~&
	    ~$1.712^{+0.318}_{-0.174}$~&~$1.377^{+0.514}_{-0.724}$~&~$1.474^{+0.615}_{-0.163}$~\\
	   
	    $\epsilon$~&~$-0.051^{+0.794}_{-0.410}$~&~$0.053^{+0.431}_{-0.306}$~&
	    ~$0.211^{+0.213}_{-0.286}$~&~$0.191^{+0.386}_{-0.236}$~&~$0.090^{+0.409}_{-0.365}$~\\
		
		$\alpha$~&~$0.134^{+0.135}_{-0.638}$~&~$0.133^{+0.155}_{-0.612}$~&
		~$0.133^{+0.095}_{-0.670}$~&~$0.133^{+0.195}_{-0.153}$~&~$0.131^{+0.140}_{-0.637}$~\\
		
		$\beta$~&~$2.903^{+0.580}_{-1.450}$~&~$2.912^{+0.968}_{-1.503}$~&
		~$2.904^{+0.433}_{-0.880}$~&~$2.910^{+0.534}_{-0.431}$~&~$2.806^{+1.938}_{-0.529}$~\\
		
		$M_\mathrm{B}^1$~&~$-18.984^{+0.803}_{-0.330}$~&~$-18.995^{+0.744}_{-0.304}$~&
		~$-18.991^{+0.646}_{-0.613}$~&~$-18.987^{+1.125}_{-0.840}$~&~$-18.979^{+0.331}_{-0.457}$~\\
		
		$\Delta_\mathrm{M}$~&~$-0.011^{+0.291}_{-0.315}$~&~$0.001^{+0.904}_{-0.976}$~&
		~$-0.001^{+0.445}_{-0.837}$~&~$0.001^{+0.669}_{-0.599}$~&~$-0.004^{+1.015}_{-0.495}$~\\
		
		$a_1$~&~$-0.213^{+0.518}_{-0.315}$~&~$-0.173^{+0.349}_{-0.262}$~&
		~$-0.183^{+0.332}_{-0.400}$~&~$-0.201^{+0.443}_{-0.257}$~&~$-0.233^{+0.297}_{-0.269}$~\\
		
		$a_2$~&~$0.018^{+0.396}_{-0.269}$~&~$0.007^{+0.258}_{-0.162}$~&
		~$0.006^{+0.219}_{-0.247}$~&~$0.003^{+0.387}_{-0.271}$~&~$0.001^{+0.231}_{-0.152}$~\\
		
		\hline
		~~&~~&&$\sigma_0$&~~&~~\\
		\hline
		$\Omega_k$~&~$-0.313^{+0.485}_{-0.308}$~&~$-0.560^{+0.880}_{-0.853}$~&
			~$-0.914^{+0.902}_{-0.194}$~&~$-0.026^{+0.329}_{-0.460}$~&~$-0.450^{+0.808}_{-1.082}$~\\
			
			$\eta$~&~$2.145^{+0.860}_{-0.366}$~&~$1.815^{+0.478}_{-0.204}$~&
			~$2.120^{+0.316}_{-0.123}$~&~$1.872^{+0.343}_{-0.276}$~&~$1.761^{+0.564}_{-0.564}$~\\
			
			$\delta$~&~$1.514^{+0.250}_{-1.130}$~&~$2.037^{+0.506}_{-0.538}$~&
			~$1.006^{+0.219}_{-0.512}$~&~$1.780^{+0.576}_{-0.465}$~&~$2.044^{+0.608}_{-0.542}$~\\
			
			$\epsilon$~&~$0.192^{+0.154}_{-0.178}$~&~$0.089^{+0.494}_{-0.558}$~&
			~$0.310^{+0.377}_{-0.248}$~&~$0.350^{+0.425}_{-0.348}$~&~$0.112^{+0.335}_{-0.539}$~\\
			
			$\alpha$~&~$0.137^{+0.017}_{-0.658}$~&~$0.133^{+0.027}_{-0.656}$~&
			~$0.127^{+0.239}_{-0.546}$~&~$0.133^{+0.041}_{-0.732}$~&~$0.133^{+0.057}_{-0.665}$~\\
			
			$\beta$~&~$2.894^{+0.207}_{-0.996}$~&~$2.911^{+1.257}_{-1.362}$~&
			~$2.955^{+0.235}_{-0.421}$~&~$2.899^{+0.403}_{-1.471}$~&~$2.912^{+1.121}_{-1.175}$~\\
			
			$M_\mathrm{B}^1$~&~$-18.940^{+0.323}_{-0.411}$~&~$-18.994^{+1.297}_{-0.594}$~&
			~$-18.954^{+0.422}_{-0.243}$~&~$-18.980^{+0.523}_{-0.415}$~&~$-18.990^{+0.314}_{-0.220}$~\\
			
			$\Delta_\mathrm{M}$~&~$-0.061^{+0.249}_{-0.356}$~&~$0.002^{+0.247}_{-0.293}$~&
			~$-0.001^{+0.376}_{-0.254}$~&~$-0.005^{+0.257}_{-0.251}$~&~$-0.005^{+0.841}_{-0.549}$~\\
			
			$a_1$~&~$-0.282^{+0.114}_{-0.283}$~&~$-0.180^{+0.232}_{-0.200}$~&
			~$-0.286^{+0.091}_{-0.292}$~&~$-0.235^{+0.369}_{-0.180}$~&~$-0.192^{+0.264}_{-0.612}$~\\
			
			$a_2$~&~$0.085^{+0.477}_{-0.141}$~&~$0.002^{+0.171}_{-0.130}$~&
			~$-0.012^{+0.214}_{-0.229}$~&~$0.010^{+0.378}_{-0.306}$~&~$0.007^{+0.403}_{-0.372}$~\\
		
		\hline\hline
	\end{tabular}
	\caption{\label{tab2}Constraints on all parameters from the JLA SNe Ia and SGL observations when the more complicated model is considered.}\vspace{0.2cm}
\end{table*}

\section{Conclusions and Discussions}\label{sec4} 
In this paper, by applying the simple distance sum rule, we obtained model-independent constraints on the spatial curvature by confronting the latest JLA SNe Ia with the largest SGL observations. Along with the spatial curvature $\Omega_k$, light-curve fitting parameters accounting for distance estimations from SNe Ia observations, polynomial coefficients, and parameters characterizing the mass distribution profile in SGL observations are simultaneously constrained in a global fitting without any priors. Graphical results for some concerned parameters (e.g., $\Omega_k$ and $f$) are shown in Figures~(\ref{fig1}-\ref{fig5}) and numerical results of constraints on all parameters are summarized in Tables~(\ref{tab1}, \ref{tab2}).

In summary, compared to the results obtained in \citet{Rasanen2015}, the precision of constraints on the spatial curvature obtained in our analysis has been nearly improved by a factor of 5 due to the increasing number of the well-measured SGL data points. However, it should be pointed out that results
consistently favor a spatially closed universe at very high confidence level when the SIE model is used to characterize the density profile of lenses. These direct estimations based on geometrical optics are significantly inconsistent with those constrained from some other popular cosmological probes in the standard $\Lambda$CDM scenario~\citep{Planck2015}. Actually, in \citet{Rasanen2015}, the preference of the closed FLRW model has already been indicated using the model-independent inference from SNe Ia and SGL observations. Of course, the spatially flat case was still survived in their analysis because of the weak constraints. Moreover, we extended our analysis by considering a more geneal model for the lens desity profile and found that, in some degree, the tension between the spatially flat case and observations has been alleviated and a spatiallu closed Universe is still slightly preferred. Finally, for the sake of comparison, we also estimated the constraints on the spatial curvature in the context of the FLRW model with dust and vacuum energy only from the SGL observations. The results are shown in Figure~\ref{fig6} which suggests that these model-dependent constraints on the spatial curvature are well consistent with what obtained from direct geometrical optics, i.e., the distance sum rule. Combining all results in this work and shown in \citet{Rasanen2015}, it appears that both model-independent constraints on the spatial curvature and those estimated in the standard $\Lambda$CDM scenario from SGL observations favor a spatially closed FLRW model. Alternatively, these results may imply that there are some other unknown systematics leading to bias in estimating the spatial curvature of the Universe from SGL observations. For instance, as suggested in our analysis, combination of SGL observations from different surveys is probably an underlying source of systematics in modeling the lenses. Therefore, a large number of SGL systems observed from the same program, e.g., the Euclid satellite and the Large Synoptic Survey Telescope (LSST), in the near future will be very helpful to clarify this issue.

\section*{Acknowledgments}
We would like to thank Jun-Qing Xia, Shuo Cao, Hai Yu, and Shuxun Tian for their helpful discussions at the initial stage of this work. This work was supported by the Ministry of Science and Technology National Basic Science Program (Project 973) under Grants No. 2014CB845806, the Strategic Priority Research Program ``The Emergence of Cosmological Structure" of the Chinese Academy of Sciences (No. XDB09000000), the National Natural Science Foundation of China under Grants Nos. 11505008, 11373014, and 11633001. X.D. is supported by the China Scholarship Council.

\end{document}